\documentclass[aps,prd,twocolumn,preprintnumbers,secnumarabic,amssymb,amsmath]{revtex4}

\usepackage{graphicx,epstopdf}
\usepackage{color}
\usepackage{psfrag}
\usepackage{amssymb}
\usepackage{bbold}

\newcommand{\sss}{\scriptscriptstyle}

\newcommand{\eq}[1]{\begin{equation} #1 \end{equation}}
\newcommand{\eqa}[1]{\begin{eqnarray} #1 \end{eqnarray}}
\newcommand{\op}{\mathcal{Q}}
\newcommand{\nn}{\nonumber}
\newcommand{\bra}[1]{\langle #1 |}
\newcommand{\ket}[1]{|#1 \rangle}
\newcommand{\av}[1]{\langle #1 \rangle}
\newcommand{\gev}{\,{\rm GeV}}
\newcommand{\eps}{\epsilon}
\newcommand{\heff}{{\cal H}_{\rm eff}}

\newcommand{\sect}[1]{\section{\hspace{-0.3cm} #1}}

\begin{document}
\preprint{ICCUB-12-318}
\preprint{ECMUB-77}
\preprint{UAB-FT-721}

\vspace{2cm}

\title{Natural SUSY and Kaon Mixing in view of recent results from Lattice QCD}

\author{Federico Mescia$^{a}$ and Javier Virto$^{b}$\\ \vspace{0.6cm}}
\affiliation{
$^{a}$ 
Dept. d'Estructura i Constituents de la Mat\`eria and Institut de Ci\`encies del Cosmos,
Universitat de Barcelona, Diagonal 647, E-08028 Barcelona, Spain,\\
$^{b}$
Universitat Aut\`onoma de Barcelona, 08193 Bellaterra, Barcelona, Spain.}

\begin{abstract}
Lattice results are available for $\Delta S=2$ matrix elements for the first time in full QCD, which improve 
considerably the status of hadronic uncertainties in $K$-$\bar K$ mixing with respect to earlier phenomenological 
studies. Using an average of the ETMC and RBC-UKQCD results, we analyze $\eps_K$ in Natural SUSY. This 
scenario arises as a consistent BSM framework after the latest results from the LHC. 
The analysis is improved with respect to previous studies including next-to-leading order matching conditions of order 
$\alpha_s^3$. We derive new bounds for SUSY mass insertions in the scenario with a light third generation and 
study the implications for squark and gluino masses, compared with direct searches at the LHC. Assuming natural 
values for the flavor violating SUSY couplings of both chiralities, we find that the sbottom must be heavier than 
3~TeV for a gluino mass up to 10~TeV. In this scenario no natural values for squark and gluino masses can satisfy 
the flavor bounds. 
\end{abstract}


\maketitle

\sect{Introduction}

Flavor physics observables related to mixing and decay of $K$, $D$ and $B$ mesons pose strong bounds on New 
Physics (NP) models. A strong constraint on the scale of NP comes from the measurement of $\eps_K$, related to 
indirect CP violation in the neutral Kaon system, which sets a lower bound on the NP scale around $\Lambda\sim 
10^4$ TeV in the presence of flavor violating couplings of ${\cal O}(1)$ \cite{CiuchiniDF2,IsidoriNir}.

In order to study flavour observables, one has to face the calculation of the matrix elements of the relevant local 
operators. In the case of $\eps_K$, the matrix elements of the full set of $\Delta S=2$ operators beyond the SM has 
been recently computed in full QCD by the ETMC and RBC-UKQCD lattice QCD collaborations \cite{ETMC,RBC}. 
These results constitute a considerable improvement with respect to previous results in the quenched approximation~
\cite{Lubicz:2008am}. The model-independent bounds on the scale of New Physics imposed predominantly by the 
operator $\op_4$ has increased almost by a factor of 3 \cite{ETMC}.

These new results can be immediately used to set constraints on Supersymmetry, putting bounds either on its flavor 
violating couplings, or on the SUSY masses, if some particular scenario is chosen for the flavor violation. A first 
analysis has been performed in Ref.~\cite{Kersten:2012ed}, where they consider the QCD running between the scale 
set by the heavy squark masses, and the lower scale set by the gluino mass (and eventually a light third generation).

After the first run of the LHC, direct SUSY searches have established relatively strong bounds on the masses of the 
squarks of the first two generations, more moderate bounds on the gluino mass, and still weaker bounds on third 
generation squarks. This circumstance is in fair connection to the spirit of Natural SUSY, where the only strongly 
interacting SUSY partners required to be light are the squarks of the third generation, and to a lesser extent, the 
gluino. This generic SUSY scenario is consistent with naturalness and with current results from direct searches at the 
LHC \cite{papucci,lodone,sundrum}.

In this letter, we study the bounds imposed by $\eps_K$ on Natural SUSY taking into account the recent lattice QCD 
results for the matrix elements, as well as NLO matching conditions for the $\Delta S=2$ Wilson coefficients. We 
begin in Section \ref{s1} reviewing briefly the relevant formulae for $\eps_K$ beyond the SM, and in Section 
\ref{s2} we combine the two different sets of lattice QCD results for the matrix elements, obtaining averaged results 
to be used in the phenomenological analysis. In Section \ref{s4} we summarize the relevant details concerning flavor 
violation in Natural SUSY, in Section \ref{s5} we study the constraints on the flavor violating couplings and in Section 
\ref{sec:imp} we study the implications on squark and gluino masses, under certain generic assumptions concerning 
the flavor violation.

\sect{Kaon mixing in the SM and beyond}
\label{s1}

The parameter $\eps_K$ is given by \cite{epsKSM1}
\eq{
\eps_K = \sin\phi_\eps e^{i\phi_\eps}
\left(
\frac{{\rm Im} M^{(6)}_{12}}{\Delta m_K^{exp}} + \rho \xi\right)
}
where $M^{(6)}_{12}$ is the short distance contribution at the charm scale. Assuming non-relativistic normalization 
for matrix elements, $M^{(6)}_{12}=\langle \bar K^0 \vert {\cal H}_{\rm eff}\vert K^0\rangle$. This short distance 
contribution can be split into the SM and NP components:
\eq{
M^{(6)}_{12} = M^{\rm SM}_{12} + M^{\rm NP}_{12}\ ,
}
where the NP contribution can be related to the SM and experimental values for $\eps_K$ through
\eq{
{\rm Im} M^{\rm NP}_{12} = \dfrac{\sqrt{2}\Delta m_K^{\rm exp}}{\kappa_{\eps}}
(|\eps_K|^{exp}-|\eps_K|^{\rm SM})\ ,
\label{M12NP}}
with 
$1/\kappa_{\eps}=1/(\sqrt{2}\sin\phi_\eps)(1-\rho\xi/(\sqrt{2}|\eps_K^{exp}|))$, namely, $
\kappa_{\eps}=0.94\pm 0.02$ \cite{epsKSM1}. For the SM value of $\eps_K$ we take
\eq{
|\eps_K|^{\rm SM}=(1.9\pm 0.3)\times 10^{-3}\ ,
}
computed in Ref.~\cite{epsKSM2} and rescaled to our value of $\kappa_\eps$ and to the more recent average of 
the B parameter $\hat B_K$ given in Ref.~\cite{lattav} (see Section \ref{s2}). 

The current experimental values $|\eps_K|^{\rm exp}$ and $\Delta m_K^{\rm exp}$ are given by\cite{pdg}
\eqa{
\Delta m_K^{\rm exp}&=&(3.483\pm 0.006)\cdot 10^{-15}\,{\rm GeV}\ ,\\
|\eps_K|^{\rm exp}&=&(2.229\pm 0.010)\cdot 10^{-3}\ ,
}
which together with Eq.~(\ref{M12NP}) imply the following bound on the NP contribution:
\eq{
{\rm Im} M^{\rm NP}_{12} =(1.7\pm 1.6)\cdot 10^{-18}\ {\rm GeV}\ .
\label{bound}}

The most general effective Hamiltonian for $K$-$\bar K$ mixing beyond the SM  is given by
\eq{
\heff=\sum_{i=1}^5 C_i(\mu) \op_i(\mu) + \sum_{i=1}^3 \widetilde C_i(\mu) \widetilde \op_i(\mu)\ ,
\label{Heff0}}
where the SUSY basis of operators is
\eqa{
\op_1&=& \bar d_\alpha \gamma_\mu P_L s_\alpha\, \bar d_\beta \gamma^\mu P_L s_\beta\nn\\
\op_2&=& \bar d_\alpha P_L s_\alpha\, \bar d_\beta P_L s_\beta\nn\\
\op_3&=& \bar d_\alpha P_L s_\beta\, \bar d_\beta P_L s_\alpha
\label{SUcSYbasis}\\
\op_4&=& \bar d_\alpha P_L s_\alpha\, \bar d_\beta P_R s_\beta\nn\\
\op_5&=& \bar d_\alpha P_L s_\beta\, \bar d_\beta P_R s_\alpha\nn
\label{Heff}}
together with the chirally-flipped operators $\widetilde \op_{1,2,3}$ obtained
from $\op_{1,2,3}$ with the substitution $L\leftrightarrow R$. The chiral
projectors are defined as $P_{L,R}=(1\mp \gamma_5)/2$.

The New Physics amplitude is then given by
\eq{M_{12}^{\rm NP}=\sum_{i} C_i^{\rm NP}(\mu) 
\bra{\overline K^0}{\op_i(\mu)}\ket{K^0}\ .}
The matrix element for the SM operator $\op_1$ is related to the bag parameter $B_K$
(in the non-relativistic convention)
\eq{
\bra{\overline K^0} \op_1(\mu)\ket{K^0}=\frac13 m_K f_K^2 B_K(\mu)\ ,
}
and the matrix elements of the operators $\op_{2,3,4,5}$ are usually normalized to $\av{\op_1}$,
defining the ratios $R_i$ as
\eq{R_i(\mu)\equiv \frac{\bra{\overline K^0} \op_i(\mu)\ket{K^0}}{\bra{\overline K^0} \op_1(\mu)\ket{K^0}}\ .
\label{Rii}}
The NP Wilson coefficients must be given in the same renormalization scheme as the matrix elements, and at 
the same renormalization scale $\mu$. Since the matching conditions are computed at the matching scale $
\Lambda$ related to the masses of the heavy particles, the Wilson coefficients must be evolved down by 
means of the Renormalization Group. The evolution matrix at NLO in QCD for $\mu=2$ GeV in the RI scheme 
is given in Ref.~\cite{Ciuchini:1998}. Taking this into account, we can write
\eqa{
M_{12}^{\rm NP}&=&
\frac13 m_K f_K^2 B_K
\Big[ \xi_{11} (\Lambda) (C_1^{\rm NP}(\Lambda)+\widetilde C_1^{\rm NP}(\Lambda))\nn\\[2mm]
&&
+\,[\xi_{22}(\Lambda) R_2 + \xi_{23}(\Lambda) R_3] (C_2^{\rm NP}(\Lambda)+\widetilde C_2^{\rm NP}
(\Lambda))\nn\\[2mm]
&&
+\,[\xi_{32}(\Lambda) R_2 + \xi_{33}(\Lambda) R_3]  (C_3^{\rm NP}(\Lambda)+\widetilde
C_3^{\rm NP}(\Lambda))\nn\\[2mm]
&&
+\,[\xi_{44}(\Lambda) R_4 + \xi_{45}(\Lambda) R_5]  C_4^{\rm NP}(\Lambda)\nn\\[2mm]
&&
+\,[\xi_{54}(\Lambda) R_4 + \xi_{55}(\Lambda) R_5]  C_5^{\rm NP}(\Lambda)\Big]\ ,
\label{M12NP2}}
where the NLO evolution coefficients $\xi_{ij}(\Lambda)$ for $\Lambda=1$ TeV in
the Landau-RI scheme are collected in Table~\ref{Tabxis}. Including NLO
matching conditions for the Wilson coefficients $C_i$, the combination $\xi_{ij}(\Lambda)
C_j(\Lambda)$ is independent of the matching scale at NLO. Eqs.~(\ref{bound})
and (\ref{M12NP2}) will be used in the following sections to study the constraints from $\eps_K$ on NP.

\begin{table}
\begin{tabular}{||c|c|c||}
\hline\hline
\multicolumn{3}{||c||}{$\Lambda=1$ TeV - RI scheme}\\
\hline\hline
$\xi_{11}(\Lambda)=0.762$ & $\xi_{22}(\Lambda)=2.544$ & $\xi_{23}(\Lambda)=-0.002$\\
\hline
$\xi_{32}(\Lambda)=-0.591$& $\xi_{33}(\Lambda)=0.390$ & $\xi_{44}(\Lambda)=4.823$\\
\hline
$\xi_{45}(\Lambda)=0.186$ & $\xi_{54}(\Lambda)=1.351$ & $\xi_{55}(\Lambda)=0.875$\\
\hline\hline
\end{tabular}
\caption{
Values from ref.~\cite{Ciuchini:1998} for the NLO $\Delta F=2$ evolution coefficients 
from $\Lambda=1$ TeV to $\mu=2$ GeV in the Landau RI scheme and in the SUSY basis.}
\label{Tabxis}
\end{table}

\sect{Review of Lattice QCD results for $\Delta S=2$ matrix elements}
\label{s2}

The bag parameter $B_K$ has been calculated in full QCD by lattice groups since 2004 \cite{refsBK}. The 
average result up to 2010 for the corresponding renormalization-independent parameter $\hat B_K$ is given 
by \cite{hep-lat/1011.4408}: $\hat B_K=0.738(20)$. Recently, new refined lattice studies have become 
available  \cite{1106.3230,1111.5698,1112.4861,1201.0706}. Here, we use  the updated world average 
of Ref.~\cite{lattav}:
\eq{\hat B_K=0.7643(97)\ .}
This leads to the following value for the B-parameter in the Landau-RI renormalization scheme:
\eq{
B_K^{\sss \rm (RI)}(2\,{\rm GeV})=0.546(7)\ .
\label{bkri}
}

This year, the ratios $R_i$ in Eq.~(\ref{Rii}) have been calculated in full QCD for the first time, by the ETMC 
and RBC-UKQCD collaborations \cite{ETMC,RBC}, with $N_f=2$ and 2+1 active flavors respectively. These 
results supersede previous ones in the quenched approximation \cite{KKlat1,KKlat2}.

\begin{figure*}
\includegraphics[width=15cm]{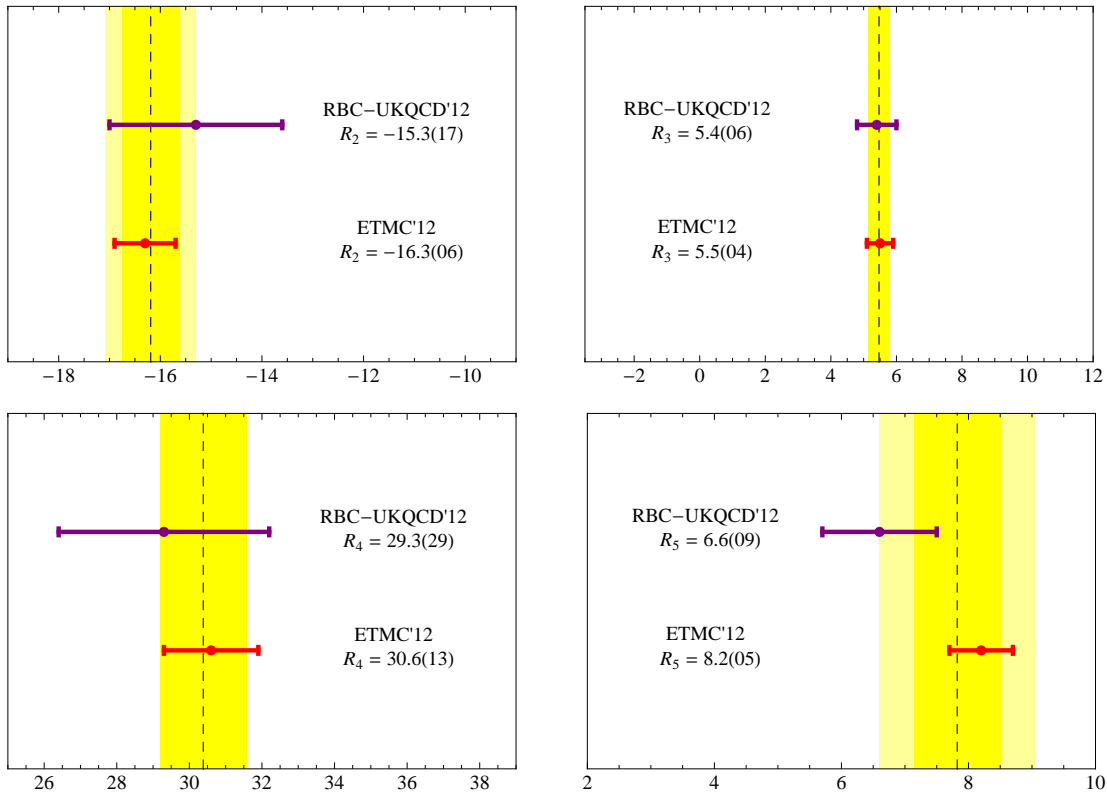}
\caption{Weighted average of unquenched lattice results of Refs~\cite{RBC,ETMC}, for the ratios of matrix 
elements $R_i$, given in the $\overline{\rm MS}$ scheme of Ref.~\cite{buras}, at the renormalization scale
$\mu=3$ GeV. The lighter colored error bars correspond to the enlarged errors that include the central values 
of the RBC-UKQCD results. See the text for details.}
\label{figlat}
\end{figure*}

The RBC-UKQCD and ETMC matrix elements are given in the SUSY basis, at a renormalization scale of 3 GeV 
and in the $\overline{\rm MS}$ scheme of Ref.~\cite{buras}. We perform a weighted average of both results 
using the procedure described in Ref.~\cite{pdg}. In the case of $R_2$ and $R_5$ the 
RBC-UKQCD central values lie outside the average error band. To take this into account we increase the errors 
to include the RBC-UKQCD central values. The RBC-UKQCD and ETMC results together with our averaged 
values are collected in Table \ref{tableRi} and represented in Fig.~\ref{figlat}.

\begin{table}
\begin{tabular}{||c||c|c|c||}
\hline\hline
\multicolumn{4}{||c||}{$\overline{\rm MS}$ at $\mu=3$ GeV}\\
\hline\hspace{1cm} & \hspace{0.8mm} ETMC \cite{ETMC} \hspace{0.8mm} & \hspace{0.8mm} RBC-UKQCD
\cite{RBC} \hspace{0.8mm} & \hspace{0.8mm} Our Average  \hspace{0.8mm} \\
\hline\hline
$R_2(\mu)$ & -16.3(0.6)& -15.3(1.7) & -16.2(0.9) \\
\hline
$R_3(\mu)$ & 5.5(0.4) & 5.4(0.6) & 5.5(0.3)\\
\hline
$R_4(\mu)$ & 30.6(1.3) & 29.3(2.9) & 30.4(1.2)\\
\hline
$R_5(\mu)$ & 8.2(0.5) & 6.6(0.9) & 7.8(1.2)\\
\hline\hline
\end{tabular}
\caption{Unquenched lattice QCD results for the ratios $R_i(\mu)$ 
as given in Refs.~\cite{ETMC,RBC}, in the $\overline{\rm MS}$ scheme 
of Ref.~\cite{buras}, given at the renormalization scale $\mu=3$ GeV. Our average results are computed
 as explained in the text.}
\label{tableRi}
\end{table}

For the phenomenological analysis ---according to Eq.~(\ref{M12NP2})---  one has to combine the  $R_i$ 
averages  in Table~\ref{tableRi} with the $\xi_{ij}(\Lambda)$ factors in Table~\ref{Tabxis}. For this purpose, 
however, the quantities $R_i$  and $\xi_{ij}(\Lambda)$ must be defined using the same renormalization 
prescription. The ratios $R_i$ in Table~\ref{tableRi}  are defined  in the $\overline{\rm MS}$ scheme of Ref.~
\cite{buras} and at $\mu=3$ GeV, while the coefficients $\xi_{ij}(\Lambda)$ in Table~\ref{Tabxis} are given 
in the Landau-RI scheme for $\mu=2$ GeV. We find it more convenient to transform the ratios $R_i$ to the 
Landau-RI scheme at $\mu=2$ GeV, e.g. to the prescription in which the coefficients $\xi_{ij}(\Lambda)$ are 
given.

We first perform the QCD running from 3 GeV down to 2 GeV. This running is performed at NLO by means of 
the two-loop anomalous dimensions given in Ref.~\cite{buras} (see also Ref.~\cite{scim}).
However, in Ref.~\cite{buras} the renormalization is carried out 
in  the so-called chiral basis of operators, $Q_i$ (see Eq.~(2.1) of Ref.~\cite{buras}). The translation between 
both bases is a fierz transformation: $Q_i^{\rm sch}(\mu)=\Delta_{ij} \op_j^{\rm sch}(\mu)$, where the 
matrix $\Delta$ is given by
\eq{
\Delta=\left(
\begin{array}{ccccc}
1&0&0&0&0\\
0&0&0&0&-2\\
0&0&0&1&0\\
0&1&0&0&0\\
0&4&8&0&0
\end{array}
\right)\ .
\label{Delta}}
where the label `sch' stands for either scheme, $\overline{\rm MS}$ or RI.
Fierz transformations introduce a different prescription for evanescent operators in the $\overline{\rm MS}$ 
scheme, which makes the $\overline{\rm MS}$ scheme of Ref.~\cite{buras} used by RBC-UKQCD and ETMC 
different from the $\overline{\rm MS}$ scheme in Refs.~\cite{scim,diego}. 

The QCD running from 3 GeV down to 2 GeV is given by
$Q^{\overline{\rm MS}}_i(2\gev) = \hat U(3\gev,2\gev)_{ji} 
Q^{\overline{\rm MS}}_j(3\gev)$, where $\hat U(3\gev,2\gev)$ is the NLO evolution matrix in the chiral 
basis and the $\rm \overline{MS}$ scheme of Ref.~\cite{buras}, given by  
\eq{
{\small \hat U(3,2)=\left(
\begin{array}{ccccc}
1.035&0&0&0&0\\
0&1.022&0.011&0&0\\
0&0.130&0.830&0&0\\
0&0&0&0.887&-0.474\\
0&0&0&0.001&1.152
\end{array}
\right)}\ .
\label{U32}}
The evolution is performed in 4-flavor QCD~\cite{0102316}, consistent with the fact that 
the charm quark is a dynamical degree of freedom from  $3$ to $2$ GeV 
for the NP contributions parametrized in Eq.~(\ref{M12NP2}) .   
The value of the strong coupling at these scales is obtained from $\alpha_s(m_c)$ running up to 2 and 3 GeV 
in the 4-flavor theory. 
We use the full results for the running of $\alpha_s(\mu)$ 
from Ref.~\cite{alphas}, giving $\alpha_s^{(4)}(2\gev)=0.3041$ 
and $\alpha_s^{(4)}(3\gev)=0.2552$. The relevant inputs at the charm scale are $\alpha_s(m_c)=0.3537$ 
and $m_c(m_c)=1.28(1)$ GeV \cite{0907.2110}.

The conversion to the RI scheme is performed 
by means of the NLO matrix $\Delta r_{\sss \rm \overline{MS}\to RI}$ of Ref.~\cite{buras}, namely 
$Q_i^{\rm RI}(\mu)=M_{ij}\,Q_i^{\rm \overline{MS}}(\mu)$, with $M=[1-(\alpha_s/4\pi) \Delta r_{\sss \rm 
\overline{MS}\to RI}]$ and $\Delta r_{\sss \rm \overline{MS}\to RI}$ given by
\eq{\hspace{1.5mm}
\Delta r_{\sss \rm \overline{MS}\to RI}=
{\small \left(
\begin{array}{ccccc}
0.879&0&0&0&0\\
0&-1.129&-6.773&0&0\\
0&0.307&10.871&0&0\\
0&0&0&5.644&0.214\\
0&0&0&12.939&2.689
\end{array}
\right)\hspace{-1cm}}
\label{MMSRI}}
This matrix can be rotated to the SUSY basis by means of the rotation $\Delta$ given in Eq.~(\ref{Delta}). 
The result will differ from the one in Refs.~\cite{scim,diego} because the $\rm \overline{MS}$ renormalization 
scheme is not the same.

Summarizing, to work out the ratios $R_i(\mu)$ at $\mu=2$ GeV in the Landau-RI scheme 
from $R_i(\mu)$ at $\mu=3$ GeV in ${\rm \overline{MS}}$ we make
\eq{
R_i^{\sss \rm (RI)}(2\gev)= {\cal N}_{ij}\,R_j^{\sss \rm (\overline{MS})}(3\gev)
}
where the transformation matrix ${\cal N}_{ij}$ is defined as
\eq{
{\cal N}_{ij}=\frac{[\Delta^{-1}M\,U^T(3,2)\Delta]_{ij}}{[\Delta^{-1}M\,U^T(3,2)\Delta]_{11}}\ .
}
Numerically, we find:
\eq{
{\cal N}=\left(
\begin{array}{c|cccc}
1&0&0&0&0\\
\hline
0&0.743&-0.037&0&0\\
0&0.073&1.083&0&0\\
0&0&0&0.608&-0.001\\
0&0&0&-0.131&1.037
\end{array}
\right)\ .
\label{Nmatrix}}
Applying this transformation to the averaged lattice results of Table \ref{tableRi}, we get
\eqa{
R_2^{\sss \rm (RI)}(2\gev)&=& -12.2(0.7)\ ,\nn\\[1mm]
R_3^{\sss \rm (RI)}(2\gev)&=& 4.8(0.3)\ ,\nn\\[1mm]
R_4^{\sss \rm (RI)}(2\gev)&=& 18.5(0.7)\ ,\\[1mm]
R_5^{\sss \rm (RI)}(2\gev)&=& 4.1(1.2)\ .\nn
\label{MERI}}
These values, together with $B_K^{\sss \rm (RI)}(2\,{\rm GeV})=0.546(7)$
of Eq.~(\ref{bkri}), will be used in the phenomenological analysis in Sections \ref{s5} and \ref{sec:imp}. 

\sect{Flavor Violation in Natural SUSY}
\label{s4}

In general SUSY models, flavor violation in the quark sector is mediated predominantly by strong interactions, 
via flavor-changing quark-squark-gluino interactions induced by soft SUSY-breaking terms. 
flavor physics.

Let ${\cal M}_q$ be the squark mass matrix in the $q=u,d$ sector, given in the super-CKM basis. In order to 
go to a physical basis where squarks do not mix with each other, a rotation is performed in the squark sector 
alone to diagonalize the squark mass matrix:
\eq{\widetilde m^2_{\rm diag} = \Gamma_q\,{\cal M}_q^2\,\Gamma_q^\dagger\ .}
After this rotation is performed, the $6\times 6$ unitary matrix $\Gamma_q$ appears in the quark-squark-
gluino vertex:
\eq{
{\cal L}_{q\tilde q \tilde g}= -\sqrt{2}g_s\Gamma_{q}^{ji*}(\tilde q^j T^a q^i)\tilde g +h.c.
}
where $q^i$ are three left-handed (i=1,2,3) and three right-handed quarks (i=4,5,6), of type $q=u$ or $d$. 
This vertex leads to squark-gluino loop penguin and box diagrams that contribute (among other things) to $
\Delta F=1$ and $\Delta F=2$ processes. As an example, the contribution to a $s\to d$ transition is given at 
the leading order by $A_{s\to d}^{\sss \rm SUSY}\sim (\alpha_s/m_{\tilde g})\Gamma^{id*}_d
\Gamma^{is}_df(\widetilde m_i^2/m_{\tilde g}^2)$, where $f(x)$ is a penguin function. It is clear that both in 
the case of degeneracy ($\widetilde m_i=\widetilde m$) and in the case of alignment ($\Gamma^{ij}=
\delta_{ij}$), the amplitude vanishes.

Mechanisms suppressing flavor violation in SUSY such as degeneracy (SUSY-GIM mechanism) or alignment 
are required by flavor physics data, if the soft SUSY-breaking scale is low to comply with naturalness. In the 
absence of such mechanisms, the NP scale must be as high as $\Lambda\gtrsim 10^4$ TeV in order to satisfy 
bounds from $K-\bar K$ mixing \cite{CiuchiniDF2,IsidoriNir} (the bounds from $B$-physics are somewhat 
weaker $\Lambda\gtrsim 10^2$ TeV). The absence of a natural symmetry-based principle providing a 
sufficiently effective suppression of flavor violation in the presence of a low SUSY scale, without challenging 
naturalness, is a manifestation of the \emph{SUSY flavor problem}.

However, naturalness does not require all the soft masses to be low, but only those linked more strongly to the 
Higgs. In the strong sector, the stops $\tilde t_{L,R}$ contribute at one loop to the higgs mass and should be 
not much heavier than about $\sim 500$ GeV, while the gluino contributes at the two-loop level and should 
not be heavier than about $\sim 1.5$ TeV \cite{papucci,lodone}, assuming that the fine tunning is not worse
than $\sim 10\%$. By $SU(2)_L$ symmetry, the ``left-handed" 
sbottom $\tilde b_L$ is also required to be light. Beyond these restrictions, first and second generation squarks 
can be heavy, providing a scale suppression to flavor violation without compromising naturalness. These type 
of SUSY models have been collected under the name of \emph{Natural SUSY}.

In Natural SUSY, the transition $s\leftrightarrow d$ mediated by first and second generation squarks is 
suppressed by their heavy masses, and the competing process where the transition is mediated by third 
generation squarks takes over, even though it is second order in flavor violating couplings. This mechanism 
relates flavor violation in $K$ and $B$ physics. $K-\bar K$ mixing sets bounds on flavor violating couplings 
related to the third family, that are comparable to those derived from $B$ physics \cite{marco,barbieri}. 
However, for this mechanism to work, the scale suppression provided by the squark masses of the first two 
generations is in general not enough, and an additional $U(2)$ flavor symmetry might be invoked 
\cite{U2sim}.

Taking into account these considerations, we consider a Natural SUSY scenario with first generation squarks of 
mass $\sim \widetilde m_h$ around $\sim 10$ TeV, and third generation squarks of mass $\sim \widetilde m_
\ell$ around $\sim 500$ GeV. A suitable parameterization of the rotation matrices $
\Gamma_{q}=(\Gamma_{q_L},\Gamma_{q_R})$ is given by \cite{marco,jv}
\eqa{
\small
\Gamma_{q_L}&=&\left(
\begin{array}{ccc}
1&0&-\hat \delta_{LL}^{q,13}\\[1mm]
0&1&-\hat \delta_{LL}^{q,23}\\[1mm]
\hat \delta_{LL}^{q,13*}c_\theta&
\hat \delta_{LL}^{q,23*}c_\theta&c_\theta\\[1mm]
0&0&0\\[1mm]
0&0&0\\[1mm]
-\hat \delta_{LL}^{q,13*}s_\theta e^{-i\phi}&
-\hat \delta_{LL}^{q,23*}s_\theta e^{-i\phi}&
-s_\theta e^{-i\phi}
\end{array}
\right)\nn\\[5mm]
\Gamma_{q_R}&=&\left(
\begin{array}{ccc}
0&0&0\\[1mm]
0&0&0\\[1mm]
\hat \delta_{RR}^{q,13*}s_\theta e^{i\phi}&
\hat \delta_{RR}^{q,23*}s_\theta e^{i\phi}&
s_\theta e^{i\phi}\\[1mm]
1&0&-\hat \delta_{RR}^{q,13}\\[1mm]
0&1&-\hat \delta_{RR}^{q,23}\\[1mm]
\hat \delta_{RR}^{q,13*}c_\theta&
\hat \delta_{RR}^{q,23*}c_\theta&c_\theta\\
\end{array}
\right)
\label{eq:Gamma}
}
where $c_\theta=\cos{\theta_q}$ and $s_\theta=\sin{\theta_q}$, with $\theta_q$ the mixing angle in the 
$q^3_{LR}$ sector. 
The mass insertions $\hat \delta_{LL,RR}^{q,i3}$ are the couplings responsible for the flavor transitions, and 
can be bounded imposing flavor constraints. A similar parameterization for rotation matrices with non-
degenerate squarks has been considered, for example, in the phenomenological analyses in Refs.~
\cite{GNK,bkk,bkk2}, an important difference being that $\delta^{db}_{LL},\delta^{db}_{RR}$ were set to 
zero to kill effects in kaon physics.

In the next section we consider the bounds that can be derived from $\epsilon_K$ assuming a  squark 
spectrum of the type discussed above. On the other hand, these mass insertions receive contributions from soft 
SUSY-breaking parameters in the Lagrangian, as well as from Yukawa couplings. Assuming no particular 
cancellation between these two (in principle unrelated) contributions, leads to a natural size of the mass 
insertions that can be used to infer bounds on squark and gluino masses. This is the target of Section 
\ref{sec:imp}.

In order to study the constraints from flavor observables, the SUSY amplitudes must be computed. The model-
dependent part of these amplitudes is encoded in the \emph{matching conditions}, that is, the values of the 
Wilson coefficients in the effective Hamiltonian at the matching scale $\Lambda$. These matching conditions 
are known to NLO in strong interactions: leading order matching conditions can be found in Refs.~
\cite{LODF1} and \cite{LODF2} for $\Delta F=1$ and $\Delta F=2$ processes respectively. Two-loop NLO 
corrections to $\Delta F=1$ have been computed in Refs.~\cite{bobeth,pilipp}, while the full NLO corrections 
to $\Delta F=2$ can be found in Refs.~\cite{jv,jv2}.

While it can be argued that NLO corrections are numerically small and  have no
real impact on the bounds derived for the SUSY parameters, it should be noted
that at leading order the amplitude suffers from a substantial renormalization
scale dependence that leads to large uncertainties. The two main reasons for
this sensitivity to the renormalization scale are \cite{jv2,diego}: (a) the
leading order contribution is proportional to $\alpha_s^2$, while there is no
definition of the renormalization point at LO, and (b) the anomalous dimensions 
of the operators in Eq.~(\ref{Heff}) are large.

\begin{figure}
\includegraphics[width=7cm]{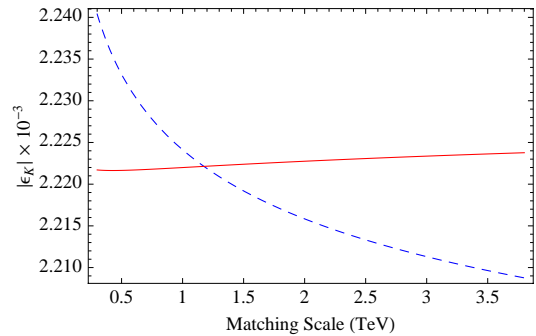}
\caption{$|\epsilon_K|$ in SM + SUSY vs the matching scale for a set of SUSY parameters consistent with 
experimental bounds. The blue dashed line is LO and the red solid line is NLO. The point at which the LO and 
the NLO coincide depends on the point in SUSY parameter space.}
\label{figScale}
\end{figure}

In order to stress this point we show, in Fig.~\ref{figScale}, the dependence of $|\epsilon_K|$ on the SUSY 
matching scale comparing the LO and NLO results. There is clearly a considerable reduction in the 
renormalization scale ambiguity when going from a LO to a NLO matching. By performing a complete NLO 
analysis, it is justified to ignore the uncertainty related to the variation of the renormalization scale. We 
emphasize that a complete NLO analysis in non-degenerate SUSY scenarios has never been done before, and 
we also note that, in general, existing LO analyses do not take into account the renormalization scale 
uncertainty.

Besides the renormalization of $\alpha_s$ and squark and gluino masses that must be taken into account at 
NLO, flavor changing renormalization of quark and squark propagators have to be considered. The (finite) 
renormalization of quark fields induced by squark-gluino loops leads to chirally-enhanced effects that can be 
numerically important (see Refs.~\cite{criv1,criv2}). However, in an ``on-shell" scheme for the super-CKM 
basis these corrections are absent. The difference between both schemes boils down to a different definition for 
the mass insertions (see appendix C of Ref.~\cite{jv2}). In this letter all mass insertions are defined in the on-
shell scheme. The (infinite) renormalization of squark fields induced by the squark tadpole implies that the 
diagonalization of the squark mass matrices must be performed at each renormalization scale. We therefore 
define the rotation matrices $\Gamma_q(\tilde \mu)$ at a fixed scale $\tilde \mu$, and include in the 
matching conditions the contribution from non-diagonal squark masses, which are of order $\tilde m_{ij}(\mu)
\sim \alpha_s \log{\tilde \mu/\mu}$. These in fact contribute to the RG equation and to the reduction of the 
renormalization scale uncertainty.

Apart from strong-interaction squark-gluino corrections, contributions from chargino-squark loops are relevant 
in certain scenarios due to the role of $A$-terms. We understand that both contributions are mostly 
uncorrelated in a general set-up, meaning that both contributions set independent bounds on SUSY (see for 
example Section 3 of Ref.~\cite{9703442}). In this letter we focus on the conclusions that can be taken from 
squark-gluino contributions alone. A study of the effect of chargino contributions is certainly worthwhile, but 
beyond the scope of this note.

\sect{Constraints from $\epsilon_K$ on flavor violating couplings}
\label{s5}

In this section we derive constraints on the insertions $\hat \delta_{LL,RR}^{db}$ and $\hat 
\delta_{LL,RR}^{sb}$ from the measurement of $\epsilon_K$. The bounds are obtained imposing the constraint 
in Eq.~(\ref{bound}) on the NP amplitude of Eq.~(\ref{M12NP2}), where the NLO matching conditions for 
the coefficients $C_i(\Lambda)$ are taken from Ref.~\cite{jv2}. The matching scale is fixed at $\Lambda=1$ 
TeV, which is justified at NLO according to the discussion in the previous section. The coefficients $C_i$ depend 
on the gluino mass $m_{\tilde g}$, the heavy and light squark masses $\widetilde m_h$, $\widetilde m_\ell$, 
and the rotation matrices  $\Gamma_q$, all defined at the matching scale. For the rotation matrices we use the
parameterization of Eq.~(\ref{eq:Gamma}). 

For the analysis we fix the masses to $\widetilde m_h=10$ TeV, $\widetilde m_{b_L}=500$ GeV, $\widetilde 
m_{b_R}=700$ GeV and $m_{\tilde g}= 1$ TeV. We put the flavor violation in the up sector to zero, and we 
consider two scenarios: $\hat \delta_{RR}^{ib}=0$ (LL only) and $\hat \delta_{LL}^{ib}=\hat 
\delta_{RR}^{ib}$ (LL=RR).

\begin{figure}
\psfrag{0.000}{\hspace{2mm} 0} \psfrag{0.00}{\hspace{2mm} 0}
\includegraphics[width=8.5cm]{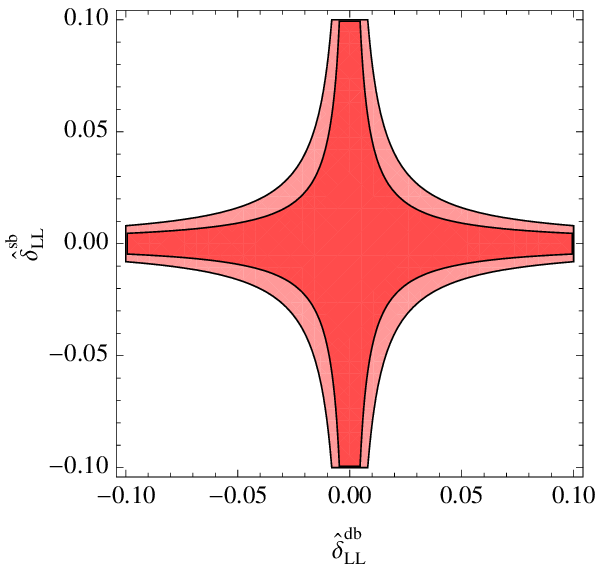}\\[4mm]
\includegraphics[width=8.5cm]{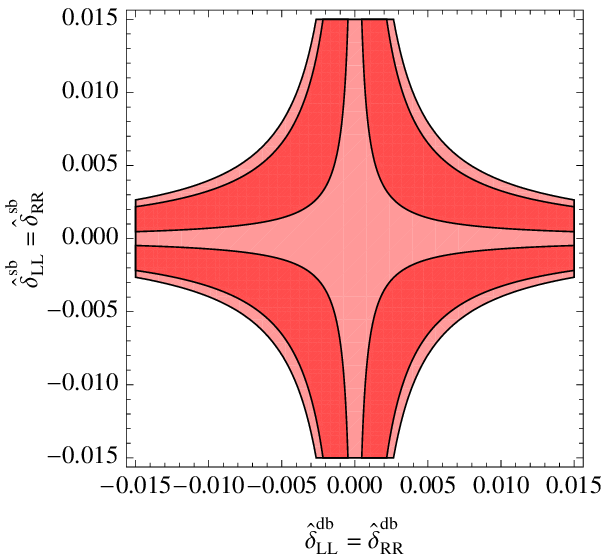}
\caption{Constraints on mass insertions from $|\epsilon_K|$ in the case of LL mixing only (upper plot) and 
LL=RR (lower plot), at one sigma (dark red) and two sigma (light red). The constraints are obtained as 
explained in the text.}
\label{figdeltas}
\end{figure}

In Fig.~\ref{figdeltas} (upper plot) we show the one and two-sigma constraints on the $\hat \delta_{LL}^{db}-
\hat \delta_{LL}^{sb}$ plane in the case of LL mixing only. The constraints are obtained by a standard 
$\chi$-square minimization, in the situation where the complex phases make the amplitude purely imaginary. 
These constraints are therefore the most conservative on the magnitude of the mass insertions,  $|\hat 
\delta_{LL}^{ib}|$. These bounds can be approximately summarized by the constraint:
\eq{
{\rm Im}\big[(\hat \delta_{LL}^{db} \hat \delta_{LL}^{sb*})^2\big]< 1.7\cdot 10^{-6}\quad {\rm at\ 95\%\, C.L.}
\label{bound1}}
The one and two sigma constraints in the case of LL=RR mixing are shown in the lower plot of Fig.~
\ref{figdeltas}. In this case, the relevant bound can be expressed approximately as
\eq{
{\rm Im}\big[\hat \delta_{LL}^{db} \hat \delta_{LL}^{sb*} \hat \delta_{RR}^{db} \hat \delta_{RR}^{sb*}\big]<
1.6\cdot 10^{-9}\quad {\rm at\ 95\%\, C.L.}
\label{bound2}}
These approximate results are obtained neglecting terms in the amplitude containing a product of more than 
four mass insertions. Since the mass insertions are small, and having checked that the numerical coefficients of 
such terms are also small, this approximation is fully justified.

\sect{Implications for squark and gluino masses}
\label{sec:imp}

\begin{figure}
\includegraphics[width=7.5cm]{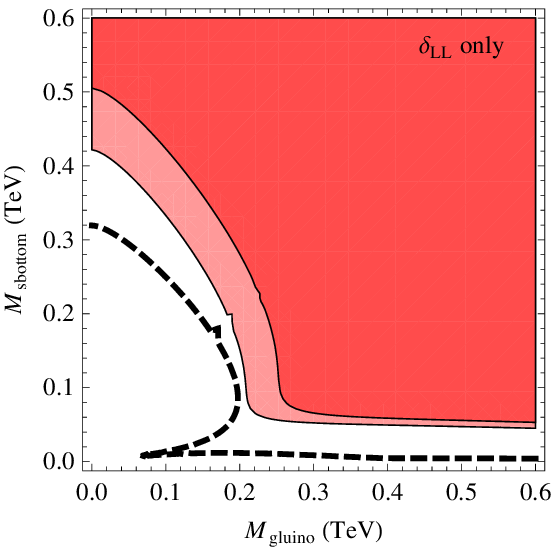}\\[4mm]
\includegraphics[width=7.5cm]{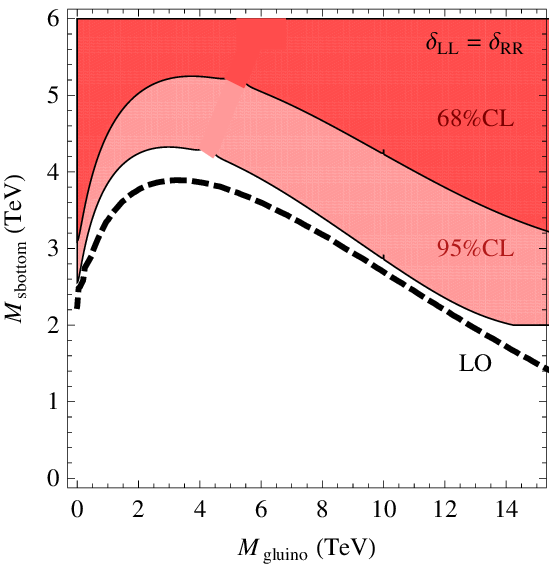}
\caption{Constraints on the masses of the gluino and the sbottom ($\tilde m_{b}=\tilde m_{b_L}=\tilde 
m_{b_R}$) from $|\epsilon_K|$ in the case of LL mixing only (up) and LL=RR (down), at 65\% (dark red) 
and 95\% C.L. (light red). The dashed line correspond to the 95\% C.L. constraint obtained from LO matching 
conditions.  The constraints are obtained as explained in the text.}
\label{figmasses}
\end{figure}

Focusing on the LL sector, the squark mass matrices in the super-CKM basis are given by ${\cal M}_u^{2\,LL} = 
V_u \tilde m^2_Q V^\dagger_u$ and 
${\cal M}_d^{2\,LL} = V_d \tilde m_Q^2 V^\dagger_d$. Here, $\tilde m_Q^{ij}$ are the soft masses for the
squark $SU(2)_L$ doublets, and the matrices $V_{u,d}$ are the rotations transforming left-handed quark 
supermultiplets from the weak to the super-CKM basis, and such that $V_u V_d^\dagger=V_{\rm CKM}$ is the 
CKM matrix. The link between ${\cal M}_u^{LL}$ and ${\cal M}_d^{LL}$ imposed by $SU(2)_L$ symmetry
is then ${\cal M}_u^{2\,LL}=V_{\rm CKM} \,{\cal M}_d^{2\,LL}\, V_{\rm CKM}^\dagger$. This relationship 
has been used to relate flavor violation in $K-\bar K$ and $D-\bar D$ mixing (see for example Refs.~
\cite{nir,criv3}). 

We can diagonalize both matrices applying the rotation matrices $\Gamma_q^{(LL)}$ in the LL sector:
\eq{\tilde m_{\rm diag}^2 = \Gamma_u\, V_{\rm CKM}\, \Gamma_d^{\dagger}\, \tilde m_{\rm diag}^2
\,  \Gamma_d\, V_{\rm CKM}^\dagger\, \Gamma_u^{\dagger}
\label{eq1}}
where $\tilde m_{\rm diag}^2=diag(\tilde m_h^2,\tilde m_h^2,\tilde m_\ell^2)$ up to perhaps terms of order 
$\tilde m_\ell^2$. We note that we have dropped the superscript $(LL)$ in the $\Gamma_q$ matrices. For 
convenience, we define ${\cal U}^\dagger=\Gamma_u\, V_{\rm CKM}\, \Gamma_d^{\dagger}$. The right 
hand side of Eq.~(\ref{eq1}) can be written as
\eq{{\rm r.h.s} =
\tilde m_h^2 \left[\mathbb{1} - {\cal U}^\dagger 
\left(
\begin{array}{ccc}
0 & &\\
& 0 &\\
& & 1\\
\end{array}
\right)
 {\cal U} + {\cal O}\bigg(\frac{\tilde m_\ell^2}{\tilde m_h^2}\bigg) \right]\ .
}
Expanding in the same way the left hand side, Eq.~(\ref{eq1}) leads to ${\cal U}_{3i}=\delta_{3i} + {\cal O}
(\tilde m_\ell^2/\tilde m_h^2)$. This equation sets a natural size for the mass insertions. For example, in the 
case in which the up quark and squark sectors are approximately aligned, we have $\Gamma_u\sim
\mathbb{1}$ and therefore $\Gamma_d^{3i} \simeq V_{\rm CKM}^{3i}+ {\cal O}(\tilde m_\ell^2/\tilde 
m_h^2)$, which translates into \cite{marco}
\eq{
\hat \delta_{LL}^{d,i3}\simeq V_{\rm CKM}^{3i}+ {\cal O}(\tilde m_\ell^2/\tilde m_h^2)\ .
\label{deltNat}}

This discussion is justified when the ratio $\tilde m_\ell^2/\tilde m_h^2$ is very small. On more general 
grounds, the condition that any chiral-conserving entry of the matrix ${\cal M}_q$ is at least of size $\tilde m_
\ell^2$, leads to $\hat \delta_{LL}\gtrsim \tilde m_\ell^2/\tilde m_h^2$ \cite{marco}.
Excluding unexpected cancellations, we expect
\eq{
\hat \delta_{LL}^{d,i3}\gtrsim {\rm max} (V_{\rm CKM}^{3i},\tilde m_\ell^2/\tilde m_h^2)\ .
}

In this section we assume that the mass insertions satisfy the lower bounds $\hat \delta_{LL,RR}^{d,i3}> V_{\rm 
CKM}^{3i}$, and study the implications of the measurement of $\eps_K$ on squark and gluino masses.

The results are shown in Fig.~\ref{figmasses} in the case of LL mixing only (upper panel) and LL=RR mixing 
(lower panel), for heavy squarks of 10 TeV and $\tilde m_{b_L}=\tilde m_{b_R}$. Also shown are the LO 
constraints, that turn out to be less stringent than the NLO ones.  In the absence of RR mixing, for a gluino 
heavier than 200 GeV, the sbottom mass is unconstrained. These bounds do not compete with direct searches 
at the LHC.
The situation is quite different in the case of LL=RR mixing (with $\tilde m_{b_L}=\tilde m_{b_R}$). In this case 
the operator $\op_4$ gives a big contribution to $\eps_K$ because of the chiral enhancement of its matrix 
element, its large anomalous dimension, and because the coefficient $C_4$ is numerically large. We find that 
the sbottom must be generically heavier than about 3 TeV independently of the gluino mass (for $m_{\tilde g}
\lesssim 10$ TeV). This situation is clearly excluded by naturalness. This is an example where the flavor bounds 
are far more stringent than the direct searches at the LHC. An intermediate scenario with $0<\delta_{RR}<\delta_{LL}$
will lead to constraints that lay in between the two extreme situations considered.

\section*{CONCLUSIONS}

We have analyzed the impact of the latest lattice QCD results for $\Delta S=2$ matrix elements in full QCD on 
Natural SUSY, with NLO matching conditions for the Wilson coefficients. The weighted average of the ETMC 
and RBC-UKQCD results for the matrix elements at 2 GeV in the Landau-RI scheme are collected in 
Eq.~(\ref{MERI}). They imply a big progress compared to older quenched results, and can be used to set 
constraints on New Physics.

Concerning the SUSY analysis, we show the impact of the inclusion of NLO matching conditions, reducing 
considerably the renormalization scale uncertainty. The bounds on the flavor violating couplings are 
summarized in Fig.~\ref{figdeltas}. They can be approximated by the bounds given in Eqs.~(\ref{bound1}) 
and (\ref{bound2}). Assuming a natural size for mass insertions (see Eq.~\ref{deltNat}), we derive lower 
bounds on squark and gluino masses. In the case of LL and RR mixing, the bounds are much stronger than the 
direct bounds from the LHC, implying a sbottom heavier than 3 TeV in this scenario.

\begin{acknowledgments}
We would like to thank M. Nardecchia for collaboration in the early stages of this work.
We are grateful  to D. Becirevic, N.~Garron, V.~Lubicz, J.~Kersten and L.~Velasco-Sevilla
for discussions. F.M. acknowledges financial support from FPA2010-20807 and the Consolider CPAN project. 
J.V. is supported in part by ICREA-Academia funds and FPA2011-25948.
\end{acknowledgments}


\begin{thebibliography}{99}


\bibitem{CiuchiniDF2} 
  M.~Bona {\it et al.}  [UTfit Collaboration],
  JHEP {\bf 0803}, 049 (2008)
  [arXiv:0707.0636 [hep-ph]].

\bibitem{IsidoriNir} 
  G.~Isidori, Y.~Nir and G.~Perez,
  Ann.\ Rev.\ Nucl.\ Part.\ Sci.\  {\bf 60}, 355 (2010)
  [arXiv:1002.0900 [hep-ph]].

\bibitem{ETMC} 
  V.~Bertone, N.~Carrasco, M.~Ciuchini, P.~Dimopoulos, R.~Frezzotti,
  V.~Gimenez, V.~Lubicz and G.~Martinelli {\it et al.},
  arXiv:1207.1287 [hep-lat].
  
\bibitem{RBC} 
  P.~A.~Boyle, N.~Garron and R.~J.~Hudspith,
  arXiv:1206.5737 [hep-lat].

  \bibitem{Lubicz:2008am} 
  V.~Lubicz and C.~Tarantino,
  Nuovo Cim.\ B {\bf 123}, 674 (2008)
  [arXiv:0807.4605 [hep-lat]].

\bibitem{Kersten:2012ed} 
  J.~Kersten and L.~Velasco-Sevilla,
  arXiv:1207.3016 [hep-ph].
  
\bibitem{papucci} 
  M.~Papucci, J.~T.~Ruderman and A.~Weiler,
  arXiv:1110.6926 [hep-ph].
  
\bibitem{lodone} 
  P.~Lodone,
  Int.\ J.\ Mod.\ Phys.\ A {\bf 27}, 1230010 (2012)
  [arXiv:1203.6227 [hep-ph]].
  
\bibitem{sundrum} 
  C.~Brust, A.~Katz, S.~Lawrence and R.~Sundrum,
  JHEP {\bf 1203}, 103 (2012)
  [arXiv:1110.6670 [hep-ph]].




\bibitem{epsKSM1} 
  A.~J.~Buras, D.~Guadagnoli and G.~Isidori,
  Phys.\ Lett.\ B {\bf 688}, 309 (2010)
  [arXiv:1002.3612 [hep-ph]]; A.~J.~Buras and D.~Guadagnoli,
  Phys.\ Rev.\ D {\bf 78}, 033005 (2008)
  [arXiv:0805.3887 [hep-ph]]

\bibitem{epsKSM2} 
  J.~Brod and M.~Gorbahn,
  Phys.\ Rev.\ Lett.\  {\bf 108}, 121801 (2012)
  [arXiv:1108.2036 [hep-ph]].

\bibitem{lattav} 
  J.~Laiho, E.~Lunghi and R.~S.~Van de Water,
  Phys.\ Rev.\ D {\bf 81}, 034503 (2010)
  [arXiv:0910.2928 [hep-ph]]. Updates: {\tt www.latticeaverages.org}.

 \bibitem{pdg} 
  J.~Beringer {\it et al.}  (Particle Data Group),
  Phys.\ Rev.\ D~{\bf 86}, 010001 (2012). 


\bibitem{Ciuchini:1998} 
  M.~Ciuchini, V.~Lubicz, L.~Conti, A.~Vladikas, A.~Donini,
  E.~Franco, G.~Martinelli and I.~Scimemi {\it et al.},
  JHEP {\bf 9810}, 008 (1998)
  [hep-ph/9808328].
 
  
\bibitem{refsBK}
Y.~Aoki,  {\it et al.},
  Phys.\ Rev.\ D {\bf 84}, 014503 (2011)
  [arXiv:1012.4178 [hep-lat]].
  M.~Constantinou {\it et al.}  [ETM Collaboration],
  Phys.\ Rev.\ D {\bf 83}, 014505 (2011)
  [arXiv:1009.5606 [hep-lat]].
  T.~Bae, {\it et al.},
  Phys.\ Rev.\ D {\bf 82}, 114509 (2010)
  [arXiv:1008.5179 [hep-lat]].
  C.~Aubin, J.~Laiho and R.~S.~Van de Water,
  Phys.\ Rev.\ D {\bf 81}, 014507 (2010)
  [arXiv:0905.3947 [hep-lat]].
  C.~Allton {\it et al.}  [RBC-UKQCD Collaboration],
  Phys.\ Rev.\ D {\bf 78}, 114509 (2008)
  [arXiv:0804.0473 [hep-lat]].
  S.~Aoki {\it et al.}  [JLQCD Collaboration],
  Phys.\ Rev.\ D {\bf 77}, 094503 (2008)
  [arXiv:0801.4186 [hep-lat]].
  D.~J.~Antonio {\it et al.}  [RBC and UKQCD Collaborations],
  Phys.\ Rev.\ Lett.\  {\bf 100}, 032001 (2008)
  [hep-ph/0702042 [HEP-PH]].
  Y.~Aoki, {\it et al.},
  Phys.\ Rev.\ D {\bf 72}, 114505 (2005)
  [hep-lat/0411006].
  J.~M.~Flynn {\it et al.}  [UKQCD Collaboration],
  JHEP {\bf 0411}, 049 (2004)
  [hep-lat/0406013].
   E.~Gamiz {\it et al.}  [HPQCD and UKQCD Collaborations],
  Phys.\ Rev.\ D {\bf 73}, 114502 (2006)
  [hep-lat/0603023].
  
\bibitem{hep-lat/1011.4408} 
  G.~Colangelo, {\it et al.},
  Eur.\ Phys.\ J.\ C {\bf 71}, 1695 (2011)
  [arXiv:1011.4408 [hep-lat]].
  
  
  
\bibitem{1106.3230} 
  S.~Durr,  {\it et al.},
  Phys.\ Lett.\ B {\bf 705}, 477 (2011)
  [arXiv:1106.3230 [hep-lat]].
  
\bibitem{1112.4861} 
  J.~Laiho and R.~S.~Van de Water,
  PoS LATTICE {\bf 2011}, 293 (2011)
  [arXiv:1112.4861 [hep-lat]].

\bibitem{1201.0706} 
  C.~Kelly,
  PoS LATTICE {\bf 2011}, 285 (2011)
  [arXiv:1201.0706 [hep-lat]].

\bibitem{1111.5698} 
  T.~Bae,  {\it et al.},
  arXiv:1111.5698 [hep-lat].


  
\bibitem{KKlat1} 
  C.~R.~Allton,  {\it et al.},
  Phys.\ Lett.\ B {\bf 453}, 30 (1999)
  [hep-lat/9806016].

\bibitem{KKlat2} 
  R.~Babich, {\it et al.},
  Phys.\ Rev.\ D {\bf 74}, 073009 (2006)
  [hep-lat/0605016].
 
 
\bibitem{alphas} 
  K.~G.~Chetyrkin, J.~H.~Kuhn and M.~Steinhauser,
  Comput.\ Phys.\ Commun.\  {\bf 133}, 43 (2000)
  [hep-ph/0004189].
 
\bibitem{buras} 
  A.~J.~Buras, M.~Misiak and J.~Urban,
  Nucl.\ Phys.\ B {\bf 586}, 397 (2000)
  [hep-ph/0005183].
  
\bibitem{scim} 
  M.~Ciuchini, 
   {\it et al.},
  Nucl.\ Phys.\ B {\bf 523}, 501 (1998)
  [hep-ph/9711402].
  
  
\bibitem{0102316} 
  A.~J.~Buras, S.~Jager and J.~Urban,
  Nucl.\ Phys.\ B {\bf 605}, 600 (2001)
  [hep-ph/0102316].

\bibitem{0907.2110} 
  K.~G.~Chetyrkin,  {\it et al.},
  Phys.\ Rev.\ D {\bf 80}, 074010 (2009)
  [arXiv:0907.2110 [hep-ph]].
  

  

  
\bibitem{barbieri} 
  R.~Barbieri, E.~Bertuzzo, M.~Farina, P.~Lodone and D.~Pappadopulo,
  JHEP {\bf 1008}, 024 (2010)
  [arXiv:1004.2256 [hep-ph]].
  


\bibitem{U2sim}
A.~Pomarol and D.~Tommasini,
  Nucl.\ Phys.\ B {\bf 466}, 3 (1996)
  [hep-ph/9507462].
  R.~Barbieri, G.~R.~Dvali and L.~J.~Hall,
  Phys.\ Lett.\ B {\bf 377}, 76 (1996)
  [hep-ph/9512388].
  R.~Barbieri, G.~Isidori, J.~Jones-Perez, P.~Lodone and D.~M.~Straub,
  Eur.\ Phys.\ J.\ C {\bf 71}, 1725 (2011)
  [arXiv:1105.2296 [hep-ph]].
  A.~Crivellin, L.~Hofer and U.~Nierste,
  arXiv:1111.0246 [hep-ph].
  R.~Barbieri, D.~Buttazzo, F.~Sala and D.~M.~Straub,
  arXiv:1203.4218 [hep-ph].
  A.~J.~Buras and J.~Girrbach,
  arXiv:1206.3878 [hep-ph].

\bibitem{marco} 
  G.~F.~Giudice, M.~Nardecchia and A.~Romanino,
  Nucl.\ Phys.\ B {\bf 813}, 156 (2009)
  [arXiv:0812.3610 [hep-ph]].

\bibitem{jv} 
  J.~Virto,
  JHEP {\bf 1201}, 120 (2012)
  [arXiv:1111.0940 [hep-ph]].
  
\bibitem{GNK} 
  Y.~Grossman, M.~Neubert and A.~L.~Kagan,
  JHEP {\bf 9910}, 029 (1999)
  [hep-ph/9909297].

\bibitem{bkk} 
  S.~Baek, D.~London, J.~Matias and J.~Virto,
  JHEP {\bf 0602}, 027 (2006)
  [hep-ph/0511295].

\bibitem{bkk2} 
  S.~Baek, D.~London, J.~Matias and J.~Virto,
  JHEP {\bf 0612}, 019 (2006)
  [hep-ph/0610109].

\bibitem{LODF1}
 S.~A.~Abel, W.~N.~Cottingham and I.~B.~Whittingham,
  Phys.\ Rev.\ D {\bf 58}, 073006 (1998)
  [hep-ph/9803401].\\
J.~M.~Gerard, W.~Grimus and A.~Raychaudhuri,
  Phys.\ Lett.\ B {\bf 145}, 400 (1984).


\bibitem{LODF2}
  J.~M.~Gerard, W.~Grimus, A.~Raychaudhuri and G.~Zoupanos,
  Phys.\ Lett.\ B {\bf 140}, 349 (1984).\\
  J.~S.~Hagelin, S.~Kelley and T.~Tanaka,
  Nucl.\ Phys.\ B {\bf 415}, 293 (1994).

\bibitem{bobeth}
C.~Bobeth, M.~Misiak and J.~Urban,
  Nucl.\ Phys.\ B {\bf 574}, 291 (2000)
  [hep-ph/9910220].

\bibitem{pilipp}
C.~Greub, T.~Hurth, V.~Pilipp, C.~Schupbach and M.~Steinhauser,
  Nucl.\ Phys.\ B {\bf 853}, 240 (2011)
  [arXiv:1105.1330 [hep-ph]].

\bibitem{jv2}
J.~Virto,
  JHEP {\bf 0911}, 055 (2009)
  [arXiv:0907.5376 [hep-ph]].
J.~Virto,
  AIP Conf.\ Proc.\  {\bf 1200}, 875 (2010)
  [arXiv:0909.3277 [hep-ph]].
  
\bibitem{diego}
M.~Ciuchini, E.~Franco, D.~Guadagnoli, V.~Lubicz, V.~Porretti and L.~Silvestrini,
  JHEP {\bf 0609}, 013 (2006)
  [hep-ph/0606197].

\bibitem{criv1} 
  A.~Crivellin and U.~Nierste,
  Phys.\ Rev.\ D {\bf 79}, 035018 (2009)
  [arXiv:0810.1613 [hep-ph]].

\bibitem{criv2} 
  A.~Crivellin and U.~Nierste,
  Phys.\ Rev.\ D {\bf 81}, 095007 (2010)
  [arXiv:0908.4404 [hep-ph]].
  
\bibitem{9703442} 
  M.~Misiak, S.~Pokorski and J.~Rosiek,
  Adv.\ Ser.\ Direct.\ High Energy Phys.\  {\bf 15}, 795 (1998)
  [hep-ph/9703442].

  \bibitem{nir} 
  K.~Blum, Y.~Grossman, Y.~Nir and G.~Perez,
  Phys.\ Rev.\ Lett.\  {\bf 102}, 211802 (2009)
  [arXiv:0903.2118 [hep-ph]].
  
\bibitem{criv3} 
  A.~Crivellin and M.~Davidkov,
  Phys.\ Rev.\ D {\bf 81}, 095004 (2010)
  [arXiv:1002.2653 [hep-ph]].
  


\end{thebibliography}
\end{document}